# Longwave infrared multispectral image sensor system using aluminum-germanium plasmonic filter arrays

Noor E Karishma Shaik[1] (✉), Bryce Widdicombe[1], Dechuan Sun[1], Sam E John[2], Dongryeol Ryu[3], Ampalavanapillai Nirmalathas[1], Ranjith R Unnithan[1] (✉)

[1] Department of Electrical and Electronic Engineering, University of Melbourne, Parkville, VIC 3010, AUSTRALIA
[2] Department of Biomedical Engineering, University of Melbourne, Parkville, VIC 3010, AUSTRALIA
[3] Department of Infrastructure Engineering, University of Melbourne, Parkville, VIC 3010, AUSTRALIA

**ABSTRACT**
A multispectral camera records image data in various wavelengths across the electromagnetic spectrum to acquire additional information that a conventional camera fails to capture. With the advent of high-resolution image sensors and colour filter technologies, multispectral imagers in the visible wavelengths have become popular with increasing commercial viability in the last decade. However, multispectral imaging in longwave infrared (LWIR; 8 – 14 µm) is still an emerging area due to the limited availability of optical materials, filter technologies, and high-resolution sensors. Images from LWIR multispectral cameras can capture emission spectra of objects to extract additional information that a human eye fails to capture and thus have important applications in precision agriculture, forestry, medicine, and object identification. In this work, we experimentally demonstrate an LWIR multispectral image sensor with three wavelength bands using optical elements made of an aluminum-based plasmonic filter array sandwiched in germanium. To realize the multispectral sensor, the filter arrays are then integrated into a 3D printed wheel stacked on a low-resolution monochrome thermal sensor. Our prototype device is calibrated using a blackbody and its thermal output has been enhanced with computer vision methods. By applying a state-of-the-art deep learning method, we have also reconstructed multispectral images to a better spatial resolution. Scientifically, our work demonstrates a versatile spectral thermography technique for detecting target signatures in the LWIR range and other advanced spectral analyses.

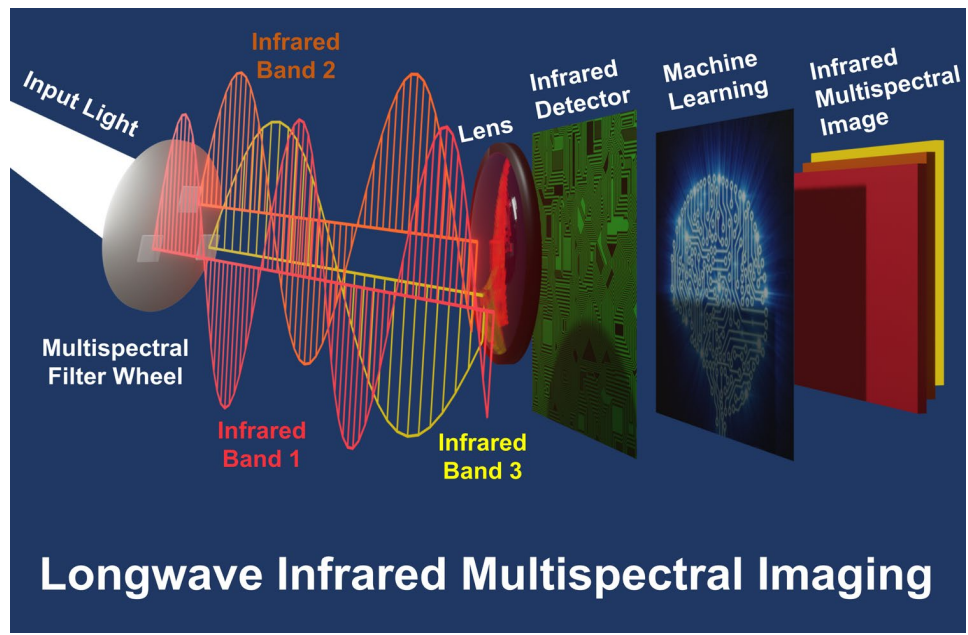

Address correspondence to Noor E Karishma Shaik, nshaik@student.unimelb.edu.au; Ranjith R Unnithan, r.ranjith@unimelb.edu.au

**1 INTRODUCTION.** Longwave infrared (LWIR) imaging cameras enable non-destructive material study and spectral analysis for various applications in defence [1], chemical detection [2], environmental sensing [3], and precision agriculture [4]. State-of-the-art thermal LWIR cameras operate in the wavelength range of 8 to 14 μm, which is transparent to low-light, fog, smoke, and other constrained environments beyond what human eyes can detect. Most such cameras contain microbolometer-based detectors that image the heat radiation from the target scene as grayscale (monochrome equivalent) pixel intensities of thermography. The grayscale thermal image sensors record an average emission spectrum of objects in the scene across a wide wavelength range and hence their spectral information is overlapped. On the contrary, multispectral imagers can capture the emission spectra in a narrow spectral band to partially resolve the emission spectra and recover additional information from the objects otherwise not possible. However, multispectral imaging is still challenging in the thermal wavelengths compared to its counterpart in the visible and near-infrared due to limited materials responding to the thermal wavelengths to make filters or filter mosaics and the low resolution of the thermal image sensors [5]. Recently, there has been a lot of interest in improving spectral resolution to enable thermal cameras as the sole imaging platform for recovering infrared fingerprints [6-8]. For multi- and hyperspectral imaging applications, state-of-the-art commercial solutions involve complex filter wheel systems or high-precision mechanical scanners that make the system expensive, bulky, and conditionally portable [9, 10]. This has led to exploring novel metasurface filters and metamaterials to make filters operating in the thermal wavelengths [11, 12].

Spoof surface plasmon (SSP) resonances occurring at the metal/dielectric interface can be exploited to make wavelength filters in the LWIR region [13]. The simplicity of the metal-dielectric structure with minimal layers and lateral tunability of its peak position makes them integrable and attractive as a two-dimensional high-density array or extended monotonic array of on-chip spectral filters for large-area applications [14-17]. Hole-based plasmonic filters integrated on CMOS chips operating in the visible range (400 – 700 nm) are reported with increased selectivity and geometrical tunability of structure [18-20]. Hole array periodicity is varied to tune the SP resonance wavelength causing spectral shifts in the corresponding spectral curves. Recently, plasmonic filters composed of periodic microhole arrays in noble metals like gold (Au), and silver (Ag) are demonstrated by exploiting SSP resonance [21-23]. Jang et al. further demonstrated spectral imaging with a square array of micron-sized holes in Au film on gallium arsenide dielectric as an SSP structure [24]. The scope of such studies has, however, been limited to developing filters or reconstructing spectral fingerprints.

In this work, we report a thermal multispectral sensor with three bands by addressing the limited response of optical material in the thermal wavelengths using aluminum-based plasmonic filters, and the resolution of the image sensors using deep learning techniques. We firstly present an aluminum (Al) infrared plasmonic filter made of a hexagonal array of holes on a germanium (Ge) substrate as an optical element to acquire thermal multispectral images. Al is the third most abundant element on earth (less expensive), which is compatible with complementary metal-oxide-semiconductor (CMOS) manufacturing methods [25]. Ge is selected as the dielectric as it is a semiconductor with a 0.66 eV bandgap that intrinsically absorbs and filters out noise in the solar wavelengths < 1.9 μm while transmitting longer wavelengths up to 16 μm [26]. Ge has a strong infrared transmission and enhances surface confinement of plasmonic resonances in Al. The three band filters operating at 10, 12 and 14 μm are designed and optimised computationally using the finite element method implemented on the COMSOL Multiphysics® Platform. The filters are fabricated using single-stage lithography and metal deposition technique. Fabricated filters are further integrated on a 3D printed wheel and then assembled to a FLIR Lepton image sensor with an 80×60-pixel resolution to make the multispectral sensor. We have demonstrated multispectral imaging using different objects after radiometric calibration, and its application by defining a function called normalized temperature variation index. We have then improved the resolution of the spectral images by four times (from 80×60 to 320×240 pixels) using a deep learning-based residual dense network. Compared to traditional high-cost multispectral imaging systems, our work represents a significant advance in the development of high-performance multispectral imaging using low-cost thermal sensors from a practical standpoint.

**2 RESULTS AND DISCUSSION.** Hybrid two-dimensional nanomaterials composed of subwavelength scale holes milled periodically in an opaque metal film have achieved optical transmission in selective wavelengths [27-29]. These plasmonic metamaterials are highly tunable in a broad wavelength range with the resonance peak and bandwidth controlled by engineering the period and diameter leading to nanophotonic devices filtering in the visible light [30-34]. There is limited literature available to extend this technology to other spectral bands specific across the LWIR region of the electromagnetic spectrum [35-39]. In this work, we develop resonant filters in the LWIR range that can be tuned laterally by engineering subwavelength geometry for integration in infrared optoelectronic systems. An aluminum plasmonic pixel of pitch size, 17 μm (that of a standard bolometer pixel size) with a hexagonal hole lattice sandwiched between Ge dielectric layers is schematically illustrated in Figure 1(a)-(b). Upon illumination, selective filtering is exhibited due to multimodal excitation and surface plasmon resonances in the metal film. Theoretical modelling of the transmission peak wavelength is given in Figure S1 in the Electronic Supplementary Material (ESM).

To comprehend the transmission mechanism in proposed plasmonic structures, the numerical simulation technique of finite element method on COMSOL Multiphysics® is used. A variable-control method is adopted to optimize structural parameters like film thickness and lateral periodicity for enhanced selectivity in the normal incidence of light. Our simulations revealed that Al plasmonic array on the Ge substrate has a broad spectrum of transmission with FWHM $\geq$ 2 μm. Adding the top layer of dielectric material (Ge) to realize dielectric-metal-dielectric (DMD) design has reduced the number of peaks and improved the transmission with FWHM $\leq$ 1.5 μm. From the obtained computational results, the metal film thickness is chosen as 50 nm which is around three times the skin depth to ensure no coupling in the top and bottom dielectric layers for preserving surface plasmon resonances at the metal-dielectric interface. The thickness of a top dielectric layer is optimized to be 300 nm for ease of fabrication. This reduces the non-uniformity in the Germanium film height over hole and metal film layers to 16%. The hexagonal arrangement of the metal hole array (MHA) increases the relative aperture area, which increases the efficiency of the device. Further, circular structures are easier to fabricate than square or other polygon-shaped holes. A hexagonal MHA with an aspect ratio (hole diameter: pitch) of 0.6 exhibits narrow linewidth and sufficient dip depth, along with having high transmission (60%), and angle-independent operation.

We systematically analyze the filter spectral response with respect to the geometrical configuration of hexagonal MHAs in terms of pitch size, angle of incidence, aspect ratio, and optical properties of constituent materials. Figure 1(e) highlights the linear shift in the peak transmission spectra as a function of pitch. Figure 1(f) shows how the angle of light incidence on the DMD filter does not affect the transmission spectra. Enhancement in the transmission peak and the FWHM with an increase in the diameter at a fixed pitch (i.e., increasing the aspect ratio) is demonstrated in Figure 1(g). Since the narrower and sharper spectral transmission filters are ideal candidates, the aspect ratio is fixed at 0.6 in this work. The selection of dielectric constituting the DMD filter is also an important factor. Figure 1(h) plots transmission peak relations with the refractive index for infrared materials with Al as a metal layer. The plots show that lower index materials are better dielectric choices with enhanced spectral sensitivity to realize the filter array.

Based on the simulation results, the filters are fabricated using standard lithography, metal deposition and lift-off processes [40]. Three plasmonic filters with pitch sizes, 2.5, 3, and 3.5 μm respectively are fabricated into a 5×5 mm$^2$ area each using a single lithography step – full fabrication details are described in the Methods section. To characterize the filters, scanning electron microscopy (SEM) imaging and spectral responsivity measurements, $S_i(\lambda)$ are performed and illustrated in Figures 1(c)-(d) respectively. Detailed scanning electron micrographs of the fabricated filters are given in Figure S2 in the ESM. Each fabricated filter will be placed in tandem with the silicon-based lens and vanadium oxide microbolometer detector array of FLIR Lepton camera (80×60 pixels, $\lambda_{range}$ ~8 – 14 μm) while imaging the target scene as demonstrated schematically in Figure 2(a). We have 3D printed a lightweight, rotating wheel and a supporting base to incorporate the monochrome camera and the three filters as illustrated in Figure 2(b). Each filter slot is a perforation extended to 6×6 mm$^2$ in the wheel, and the multispectral camera measures 8×6 cm$^2$ in size. More snapshots of the real device along with detailed measurements are shown in Figures 2(c)-(e).

We utilized the optical system to acquire radiometric data of the target scene in three LWIR wavelengths, 10, 12, and 14 μm by rotating the wheel to mount each filter sequentially. The spectral radiance flux, $F_{xy}$ for the signal acquired when light, I of certain temperature, (t) impinges on the optical system with filter responsivity, $S(\lambda)$ and detector responsivity $D(\lambda)$ is given by

$$F_{xy}(i,t) = n_0 + n_1 \int_{\lambda_1}^{\lambda_n} I(t) \cdot S_i(\lambda) \cdot D(\lambda) \, d\lambda$$

where $x \in (1, 80)$ and $y \in (1, 60)$. The input signal is received from a selective band due to the filtering effect while the noise, $n_0$ is captured across the $\lambda_{range}$ of the microbolometer detector. Surface temperature gradients and filter offsets corrupt the target measurements by propagating to the parameters of interest. Further, Ge transmittance is non-monotonic midway through the LWIR at λ~11.5 μm – which could complicate radiometric measurements. Hence, target images are corrected for non-uniformity at each pixel using a dark reference frame to eliminate spectral noises by filter and optics. Since the FLIR Lepton camera's 80×60-pixel resolution is already suffering from pixelation and imaging artifacts, any further degradation in the signal quality due to the filter will be another challenge. To improve the signal reconstruction capability and system sensitivity, we performed radiometric characterization using a blackbody instrument (by LumaSense technologies) to determine the filter's imaging capability. The radiometric temperatures were measured from the blackbody source in the temperature range of 50-200 ºC using the LWIR multispectral camera. A calibration curve is generated to map the filter-acquired temperatures with reference to the source temperature as shown in Figure 2(f). Corresponding monochrome and multispectral images of blackbody are displayed in Figures 2(g) and 2(h)-(i) respectively.

Imaging experiments are performed using our multispectral imaging system prototype (Figure 3). Four spectral images of a tree are taken with a cloudy background in monochrome, band-1, band-2, and band-3 modes of the filter wheel as displayed in Figures 3(a)-(d) respectively. The resulting multispectral image acquired by combining three bands plotted in Figure 3(e) illustrate finer details as compared to the conventional monochrome image. In addition, a new differencing metric function is defined, Normalized Temperature Variation Index (NTVI) analogous to Normalized Difference Vegetation Index (NDVI) in near-infrared counterparts to measure temperature variations and extract some additional information from the spectral images.

$$Normalized\ Temperature\ Variation\ Index\ [NTVI\ (Band\ x, Band\ y)] = \frac{Band\ x - Band\ y}{Band\ x + Band\ y}$$

To address the low resolution of thermal image sensors, we demonstrate deep learning-based radiometric superresolution to improve the performance of the optical system. A deep learning-based residual dense network is implemented to increase the image resolution. As there are not enough thermal image datasets and the few available ones are in low resolution, a fusion network pre-trained to learn how to get a high-resolution RGB image from an image with a lower resolution is adopted to learn from thermal data for faster and better superresolution. This is consistent with the studies by Choi et al. [41] that demonstrated RGB-guided networks to show good enhancement. Figure 4 shows a visual comparison of the results using the superresolution algorithm. See Figure S5 in the ESM for the representative images acquired using a visible camera. The results of superresolution images in three bands were evaluated using a no-reference image metric, Naturalness Image Quality Evaluator (NIQE) [42]. The images illustrate that algorithmic enhancement of the resolution contributes to the recovery of fine details by improving the signal quality. Note that human features like nose and fingers are better distinguishable in the superresolution image of a human holding a coffee. We applied our algorithm to the multispectral image of an in-house electronic nose with nine gas sensors integrated on a PCB board (Figure 4). Superresolution images of the electronic nose enhanced the visibility to identify the discrete sensors that are heated to a constant temperature. These images have a slight specular noise even after flatfield correction due to high Ge reflectivity at room temperature. As the temperature increases, specular noise is suppressed, and edges are reconstructed efficiently with less pixelation. The superresolution efficiency could be further increased to retrieve more accurate details by introducing the RGB information using a visible camera alongside a thermal multispectral camera, or by acquiring the target scene in time-series continuously with a thermal spectral camera at the expense of frame rate.

Taken together, the deep learning enabled LWIR multispectral imager with plasmonic MHA and the monochrome camera is a practical and cost-efficient solution. In addition, we were able to use the camera to acquire the fine spectral features of a target in the scene which weren't visible with a monochrome camera. Therefore, we have provided proof of concept for the LWIR multispectral imager with plasmonic filters. These plasmonic filters with resonant hexagonal MHAs can be easily fabricated in a fast, up-scalable method with one patterning-coating cycle followed by metal-dielectric deposition for any number of spectral bands. Our fabrication-efficient design reduces complexity and scalability costs to suit imaging and other commercial large-area applications. Our design principle is scalable to multiple wavelengths in the thermal LWIR range or other regions of the electromagnetic spectrum with atmospheric transmittance to enable non-destructive spectral analysis. As mentioned previously, realizing plasmonic nanostructures with increased thermal sensitivity can improve wavelength selectivity. This could be achieved with a low-index dielectric layer like Barium Fluoride or Potassium Bromide instead of Germanium, or other metal films like Copper and Nickel. Spatial multiplexing of several spectral filters into a single filter mosaic by varying pitch size (lateral periodicity) can further increase the spectral resolution; several joint demosaicing and superresolution algorithms can be implemented to recover missing information with finer details. Nanooptoelectronics and deep learning can thus be combined to recover rich thermal spectral information. More importantly, such a low-cost hyperspectral thermal LWIR camera enables advanced imaging and spectroscopy applications.

3 CONCLUSIONS. In this work, we have demonstrated plasmonic imaging filters combined with a low-cost uncooled monochrome thermal sensor (17 μm pixel technology) and associated deep imaging algorithm that can make a multispectral thermal sensor system to achieve the optical performance required for advanced real-world applications. The filter design exploited an aluminum layer perforated with micron size holes with a germanium cap fabricated on a germanium substrate for achieving three different wavelengths with a spectral width of around 1.5 μm and transmission efficiency of around 60%. Further, the designed optical setup is calibrated using a blackbody and its thermal output has been enhanced using computer vision algorithms to resolve blurry and pixelated edges. The spatial resolution of the spectral images is increased to 320×240 from 80×60 using a deep learning-based residual dense network, where the performance is characterised by NIQE. Different features are extracted from the multispectral images by defining a function called NTVI. However, we believe that the significance of this work goes beyond immediate practical ones, and could be the basis for further, meaningful explorations. The proposed imaging methodology could be extended for infrared sensing applications such as detecting overshooting cloud tops. The unique feature of tuning multiple wavelengths with plasmonics by engineering sub-wavelength geometries on the same metasurface is promising to develop integrated imaging and sensing solutions in the LWIR range. We, therefore, envision that this work will spur further exploration of non-traditional optical design pathways for novel, high-performance, and low-cost infrared imaging solutions.

**MATERIALS AND METHODS**

**Simulation:** A single unit cell of the period array under study has been modelled in the x-y plane and consistently illuminated in the z-direction through a broadband light source with electric polarization along the x-direction. The top view of the simulation model is presented in Figure 1(a). The single-unit simulation model represented in Figure 1(b) consists of the Aluminum layer perforated with holes of varied thickness on a 500 nm-thick, Germanium substrate with a refractive index of 4. The refractive indices of Aluminum for varied wavelengths are taken from Rakic [43]. A Germanium layer is added on the top with holes filling the Germanium to achieve the dielectric-metal-dielectric (DMD) structure. A 500 nm air layer is added to the top of the Germanium layer. The symmetric boundary conditions with perfectly matched top and bottom layers are applied, and finer mesh with a step size of 5 nm is utilized at hole level in the single-unit cell design. The computational methods show that the magnetic field component of incoming radiation generates anti-parallel currents in the top and bottom metal, creating strong field confinement at resonance. Ideally, the transmission minimizes ohmic losses in metal and dielectric losses in the spacer. When the correct mode is excited, the angle of incidence does not shift the spectral location of resonance. Transmission parameters are plotted for multispectral filter arrays in Figure 1(e)-(f).

**Fabrication:** Three Ge-Al-Ge filters with MHA structures are fabricated in a 5×5 mm$^2$ pattern for each band using maskless lithography, multilayer deposition, and lift-off processes (See Figure S2 in the ESM for fabrication layout). To start with, 0.5 mm-thick Ge substrates of 1×1 cm$^2$ area are wetcleaned with acetone, and isopropanol (IPA) for 120 s and rinsed in distilled water (dH$_2$O) for 15 s. Samples are then dried through hard nitrogen (N$_2$) blow-off at room temperature. Clean Ge samples are spin-coated with AZ1512HS photoresist at 3500 rpm for 90 s and baked on a hot plate at 110 ºC for 60 s. Maskless lithography is performed through UV light exposure using the Direct Write Lithography tool to precisely pattern the samples with holes of 1.8-2.4 μm at a pitch in the range of 3-4 μm respectively (aspect ratio: 0.6). The samples are then developed in diluted AZ726 solution (AZ726:dH$_2$O = 2:1) followed by dH$_2$O rinse and N$_2$ hard blow. A 50 nm-thick Aluminum thin film is deposited along with a 5 nm cap layer of Germanium using ebeam evaporation for metallization and then lifted off using acetone, IPA, and dH$_2$O by drying in N$_2$ flow to form MHAs. The cap layer serves as a protective coating and ensures no rapid oxidization of Al during the lift-off process. An additional 300 nm Ge is deposited to realize a DMD array for narrow FWHM at the cost of reduced transmission. The SEM images for the filters realized are included in Figure S2 in the ESM.

**FTIR Measurement:** The transmittance spectra of the fabricated filter array were measured in the wavenumber range of 650-2000 cm$^{-1}$ (5 – 14 μm) using the Fourier-transform infrared (FTIR) spectroscopy with Thermo Fisher Scientific Nicolet is50 instrument. The sampling area was adjusted to 100×100 μm$^2$ using a variable aperture and the relative transmittance spectra for each filter were obtained by normalizing the spectra with an air background. The FTIR spectral measurements of the fabricated array plotted in Figure 1(d) have recorded an FWHM of 1.35 μm up to 60% peak transmittance.

**Imaging Experiment:** In this work, multispectral thermal images are obtained using a Lepton camera sequentially using the filters appended to a 3D printed wheel in front of the camera. The dataset is created with capturing various temperature-sensitive objects in indoor and outdoor environments. A dark reference frame is acquired after every sample for specular denoising. The recorded images were loaded in an in-house MATLAB (v2019b) script. Flatfield correction is performed using the target and reference frames to eliminate unwanted noise. The transmission intensities were extracted correspondingly at each pixel for all the filters. Radiometric calibration is performed by measuring a fixed set of temperature values ranging from 25-200 °C with LumaSense's Mikron® infrared blackbody source, and a linear intensity response behaviour was observed. This optical setup can be used to measure the temperature of various objects with spectral filters and to predict real temperature from the calibration curve.

**Superresolution Algorithm:** To solve the thermal superresolution problem without ground truth, an enhanced superresolution generative adversarial network (ESRGAN) pretrained on the visible dataset is adopted in our work [44]. A generative adversarial network encourages the network to favour solutions that look more like natural images. By using the features before activation, the perceptual loss is improved which could provide stronger supervision for brightness consistency and texture recovery. Benefitting from these improvements, the ESRGAN model is proven to achieve consistently better visual quality with more realistic and natural textures. See Figure S4 in the ESM for the complete architecture. For training, we mainly use the DIV2K dataset [45], which is a high-quality (2K resolution) dataset of 800 images with rich and diverse textures for image restoration tasks. The mini-batch size is set to 16 while training. The spatial size of cropped HR patch is 128×128. We train the model in RGB channels and augment the test dataset with RGB and thermal multispectral images. The pretrained ESRGAN has been able to produce sharper human features and enose textures.

## Acknowledgements

The numerical simulations were undertaken in the NCI National Facility in Canberra, Australia, which is supported by the Australian Commonwealth Government. This work was performed in part at the Melbourne Centre for Nanofabrication (MCN) in the Victorian Node of the Australian National Fabrication Facility (ANFF). This project received funding from the Linkage Grant from Australian Research Council (LP160101475). The authors thank Mr. Luke Weston, Dr. Nandakishor Desai, and Assoc. Prof. Karim Seghouane for helpful discussions. NKS would like to additionally acknowledge the financial support provided by Melbourne Research Scholarship during PhD.

**Electronic Supplementary Material**: Supplementary material (please give brief details, e.g., further details of the annealing and oxidation procedures, STM measurements, AFM imaging and Raman spectroscopy measurements) is available in the online version of this article at http://dx.doi.org/10.1007/s12274-***-****-* (automatically inserted by the publisher).

# FIGURES

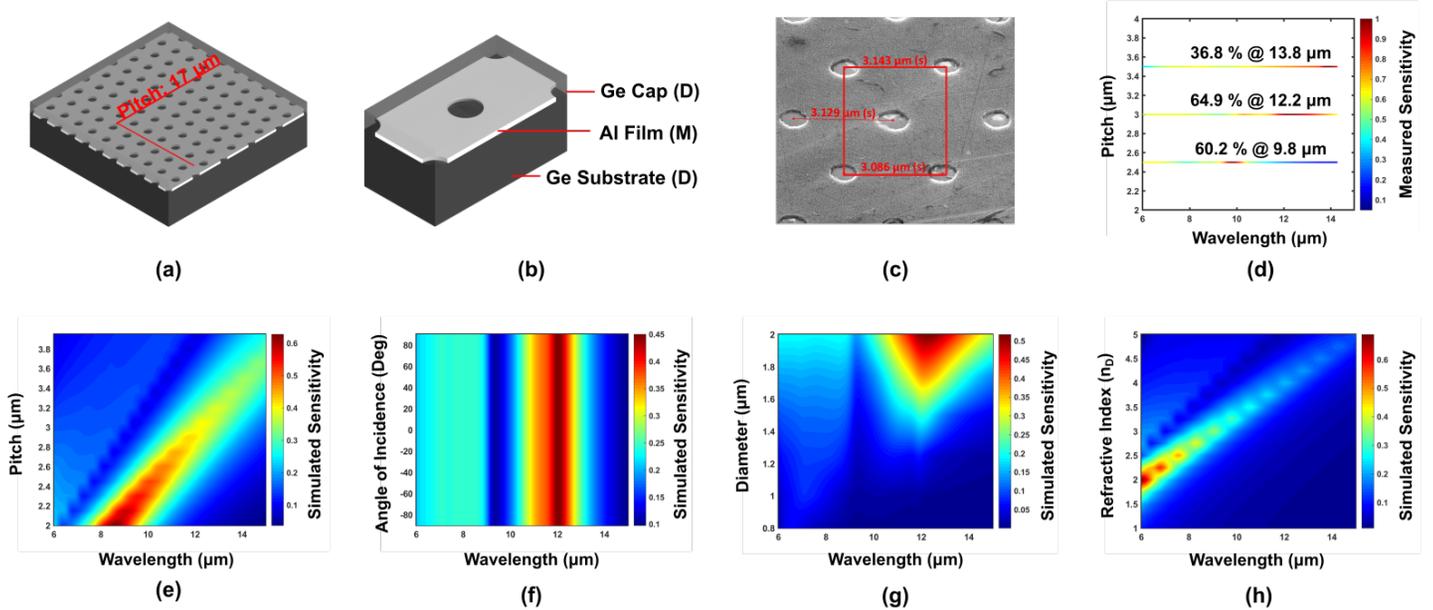

**Figure 1 Design and Fabrication of Ge-Al-Ge Plasmonic Filter Array for Multispectral Imaging:** (a) 3D schematic of a plasmonic filter array (constant pitch) covering pixel unit of size, 17 μm in a thermal camera highlighted in red. (b) A single hexagonal element of micron-sized hole lattice with 50 nm metal film sandwiched between 300 nm Ge cap and 0.5 mm Ge substrate layers to form a dielectric – metal – dielectric (DMD) filter. (c) Scanning electron micrograph of a fabricated filter with pitch – 3 μm. Scale bar – 2 μm. Aspect ratio (hole diameter: pitch size) – 0.6. For detailed SEM images, see Figure S2 in the ESM. (d) Experimentally measured transmissivity (normalized) of fabricated filters with pitch sizes: 2.5, 3 and 3.5 μm respectively. (e) – (h) present COMSOL simulation results. Filter transmission is computationally modelled for the realized DMD designs with pitch sizes in an extended range of 2 – 4 μm as shown in (e). For a filter with the fixed pitch of 3 μm, (f) and (g) demonstrate the spectral insensitivity to the angle of incidence and sensitivity to hole diameter respectively. Our simulations further illustrate that tuning the refractive index of the dielectric layer (D) by choosing an infrared material with a lower refractive index (like Barium Fluoride or Potassium Bromide etc..) can enable improved spectral sensitivity as shown in (h) in future designs.

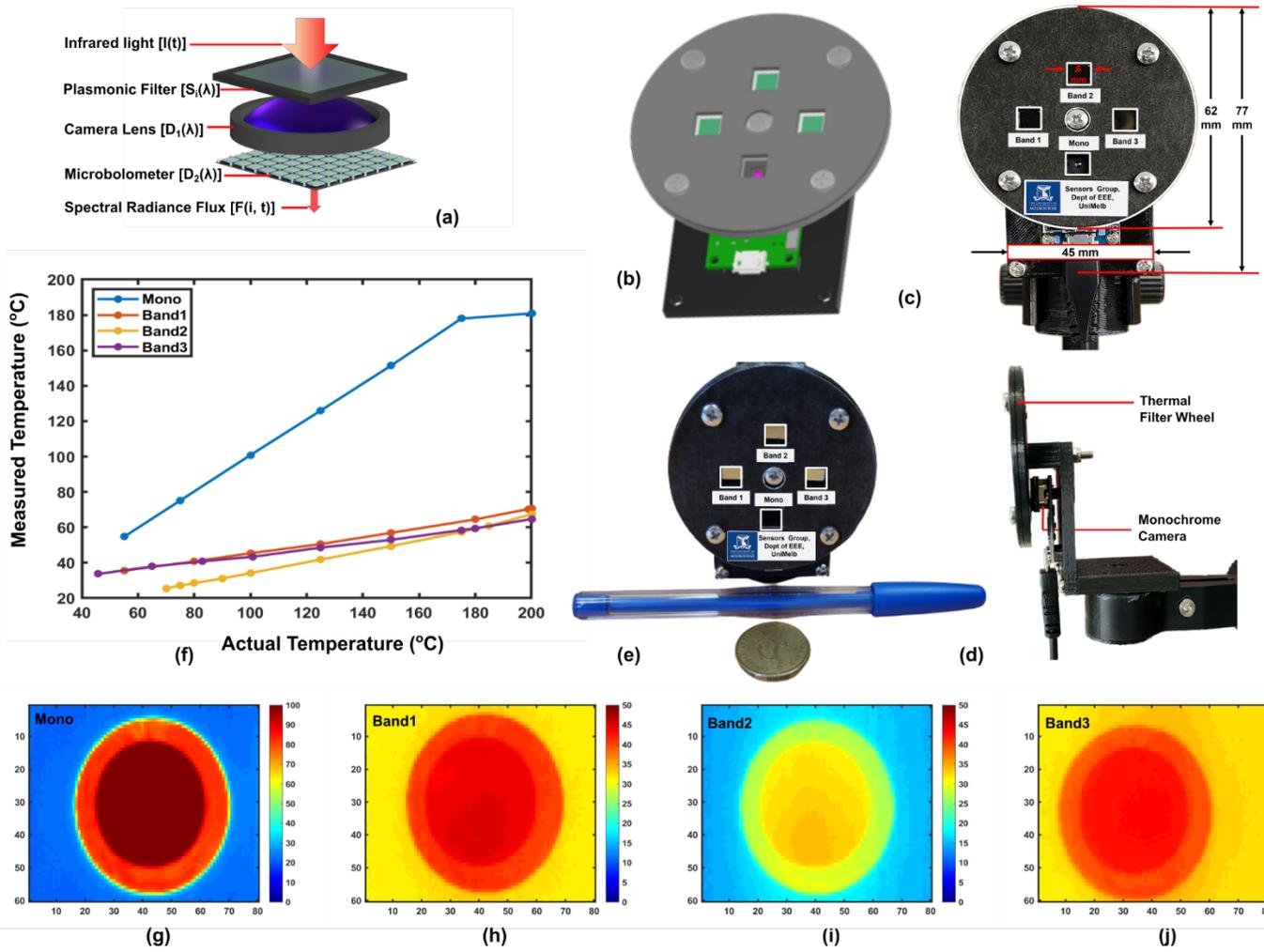

**Figure 2 Thermal Multispectral Camera Integration and Characterization:** (a) Optical path for light manipulation, from top to bottom: incident infrared light, fabricated plasmonic filter, camera lens appended to microbolometer [with detector responsivity $D(\lambda) = D_1(\lambda) \cdot D_2(\lambda)$] and filtered light. (b) Schematic for thermal multispectral wheel integrated into the commercial camera. See Figure S3 in the ESM for the exploded schematic view. (c) – (e) 3D printed filter wheel with three spectral filters (5×5 mm$^2$ each) integrated into the thermal monochrome camera visualized from the front in (c) and side in (d). The size of our proposed camera in comparison to a coin and a pen is displayed in (e). (f) – (j) Blackbody calibration curve is plotted for the camera operating temperature range across all bands in (f). Multispectral image data of blackbody at 100 ºC temperature is displayed without filter (Mono) in (g) and with three spectral filters (Band 1–3) in (h), (i), (j) respectively.

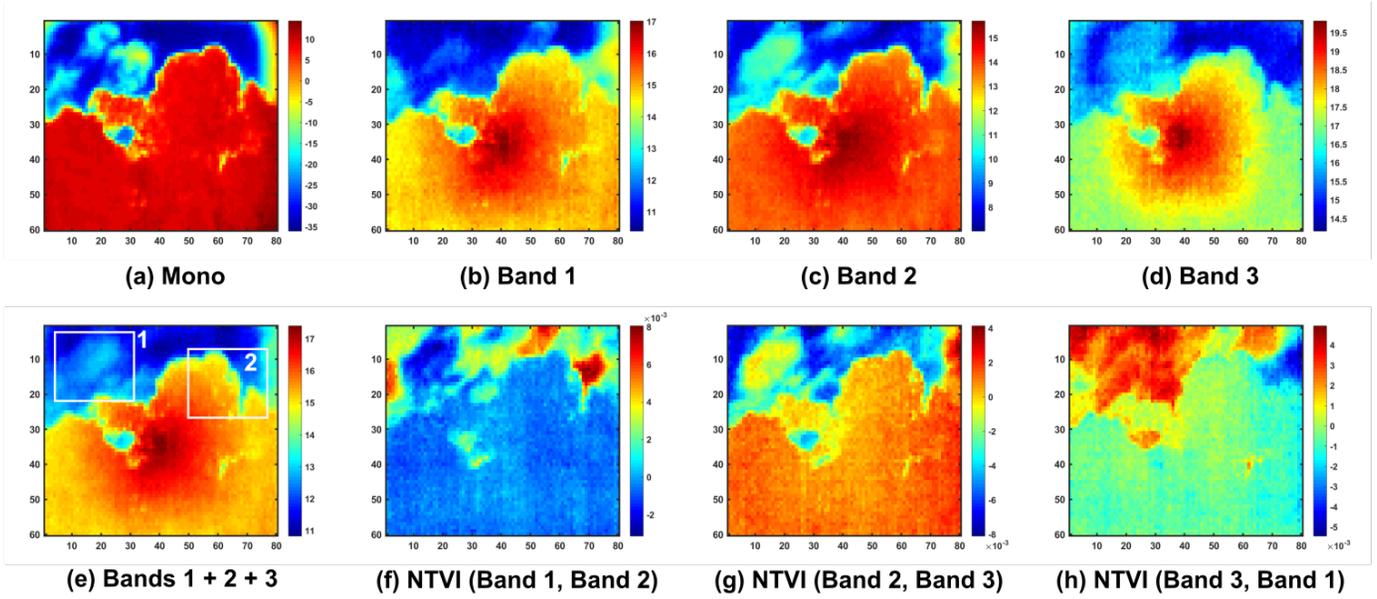

**Figure 3 Visual Results of Thermal LWIR Multispectral Imaging:** (a) – (e) Monochrome, multispectral (bands1 – 3) and fusion images of a tree with sky background are given in (a), (b) – (d) and (e) respectively. Clouds and sky are highlighted in 1 and 2 regions respectively. (f) – (h) Normalized temperature variation index measured between two thermal bands.

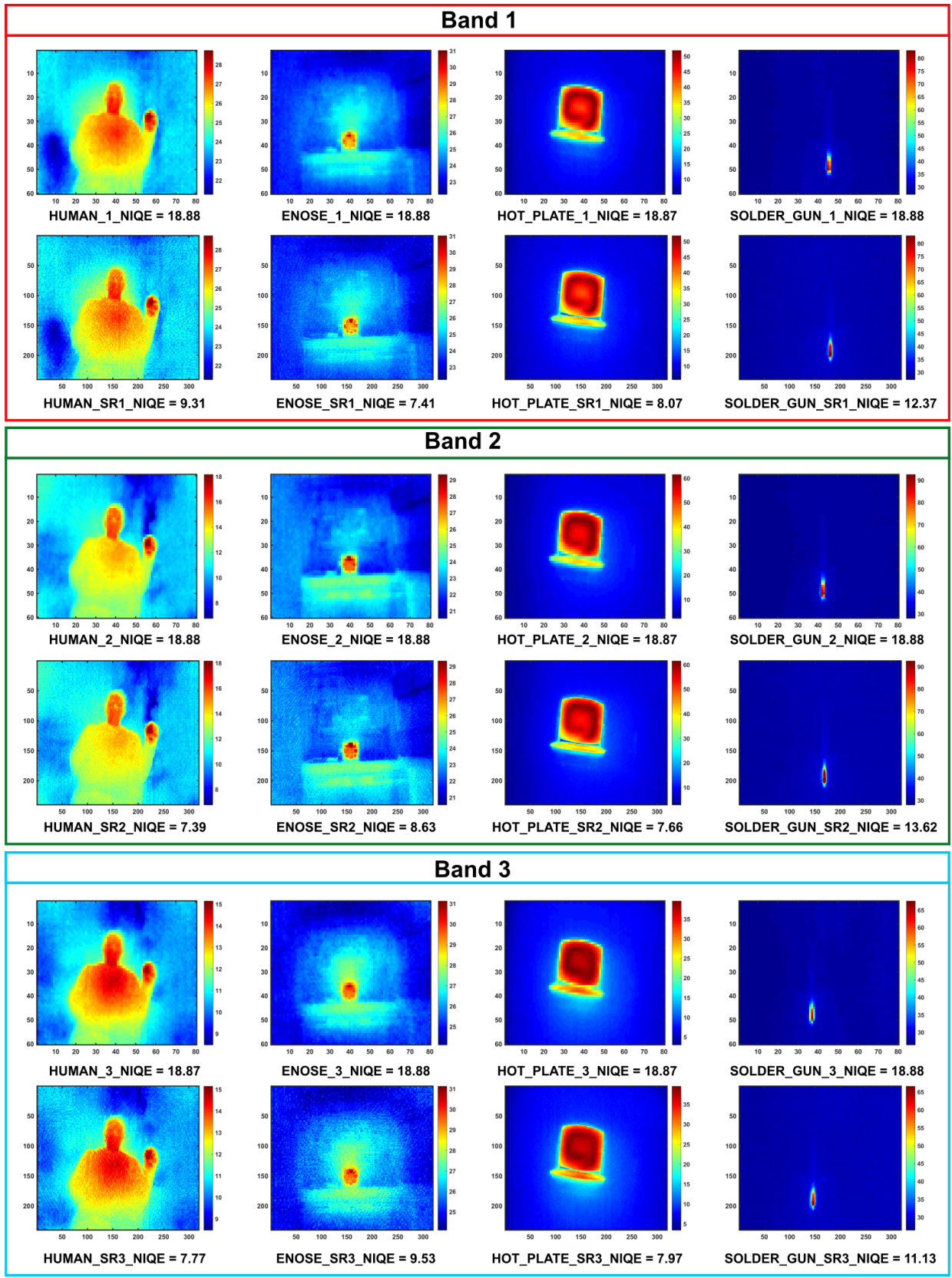

**Figure 4 Visual Results of Superresolution Imaging:** From top to bottom: input and output images for bands 1-3 respectively. Each image is the spectral data arranged in a pixel array with color bar representing the false temperature map (ºC). From left to right: human holding a coffee (30 ºC), electronic nose (40 ºC), hot toaster (150 ºC) and soldering gun (200 ºC). Visible counterparts are shown in Figure S5 in the ESM. The superresolution performance is characterized by a no-reference image quality score, Naturalness Image Quality Evaluator (NIQE). The reconstructed images have a smaller score, showing better perceptual quality.

# ELECTRONIC SUPPLEMENTARY MATERIAL

## 1. Geometrical tuning of plasmonic microhole array

The infrared transmission mechanisms in thin, opaque metal films bearing tiny holes, with sizes smaller than the wavelength of the incident light, could be used in the transmission layer to strongly enhance transmission through the holes and wavelength filtering. The hole-array-based filters operating in the transmission mode will have resonant peaks dominated by the surface plasmon mode at the microhole (cylindrical) boundaries and two surfaces. The relation describing the resonant wavelength associated with triangular or hexagonal hole array of the surface plasmons for normal incidence of light at a metal-dielectric interface at short wavelengths is given by:

$$\lambda_{SPP} = \frac{P_{SPP}}{\sqrt{\frac{4}{3}(i^2 + j^2 + ij)}} \sqrt{\frac{\epsilon_M \epsilon_D}{\epsilon_M + \epsilon_D}} \quad \ldots(1)$$

where $\lambda_{max}$ is the peak wavelength, PSPP is the pitch or period of the hole array, i and j denote the scattering orders of the array, $\epsilon_M$ is the permittivity of the metal and $\epsilon_D$ is the permittivity of the dielectric with the scattering orders i = 1 and j = 0 for the first transmission peak. At the longer wavelengths where $\epsilon_M \gg \epsilon_D$, the equation (1) can be approximated as:

$$\lambda_{SPP} = \frac{\sqrt{3} \, P_{SPP} \, n_D}{2} \quad \ldots(2)$$

where $n_D$ represents the refractive index of the dielectric layer.

Following equation 2, the peak positions are plotted for various metals at the metal-dielectric interface as given in Figure S1. According to equations 1 and 2, the peak wavelength is determined using the periodicity of the hole array in the metal film. However, the effects of hole size (relative aperture area) and metal film thickness are not included. Hence, using this equation, it is possible to learn the approximate peak position at any wavelength of interest. Infrared materials like Gallium Arsenide (GaAs, $n_D$ = 3.3) and Germanium (Ge, $n_D$ = 4) have enhanced transmission peaks above 7 μm with an increase in hole size whereas Silica glass (SiO$_2$, $n_D$ = 1.47) doesn't transmit beyond the near-infrared (NIR) region.

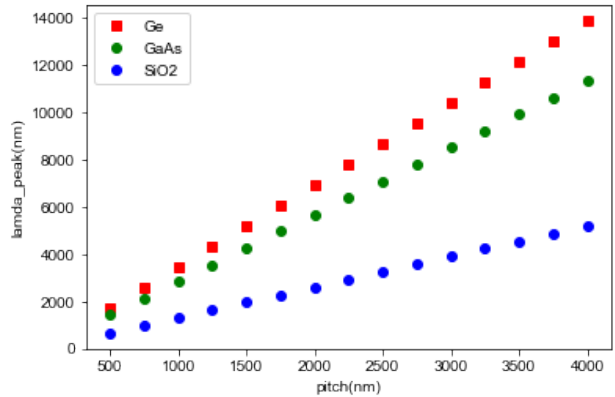

**Figure S1** Peak wavelength at longer wavelengths for various dielectric materials with Aluminum (Al) metal

There are a variety of materials that are reflective in the infrared wavelengths, as shown in Figure S1a. Gold and Silver have been the traditional choice of metals to design surface plasmon sensors with periodic nanohole arrays. In the current work, Aluminum is chosen as an alternative plasmonic material to maintain the compatibility with standard CMOS process. The additional advantages like low cost, natural abundance, and ease of processing make our choice promising for future study.

Address correspondence to Noore Karishma Shaik, nshaik@student.unimelb.edu.au; Ranjith R Unnithan, r.ranjith@unimelb.edu.au 2

## 2. SEM Characterization of fabricated microhole arrays

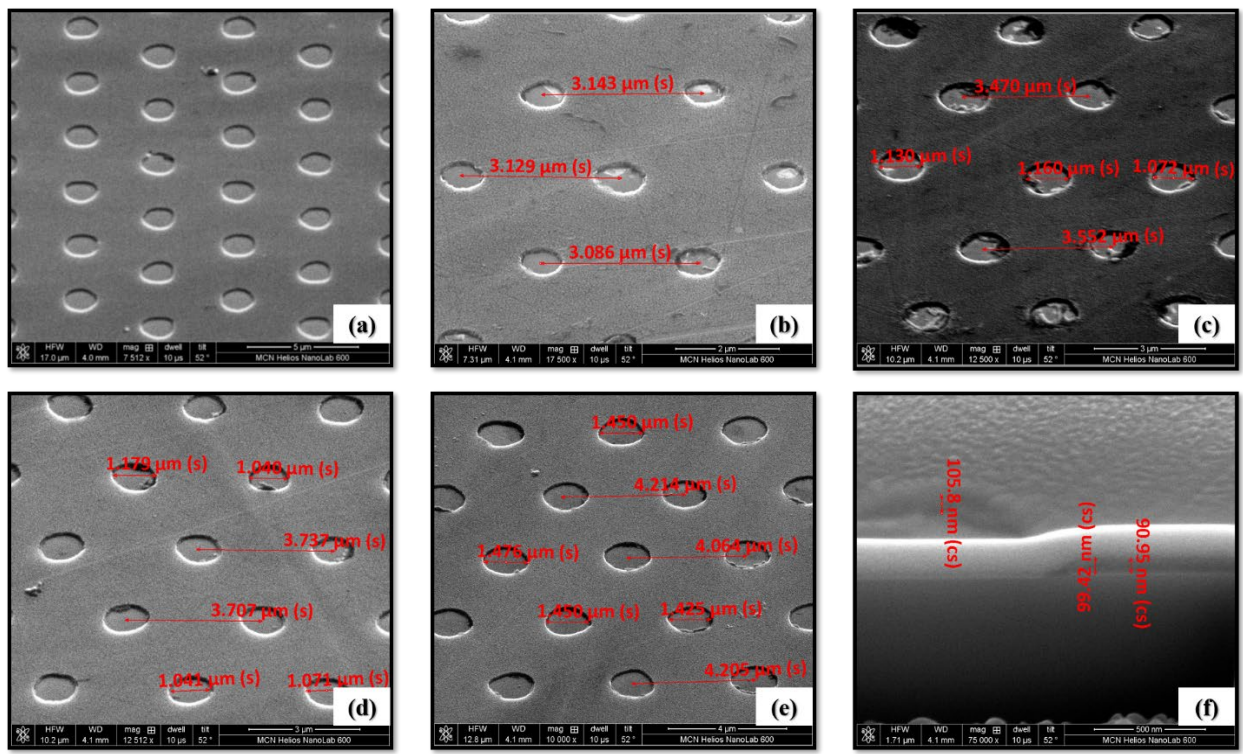

**Figure S2** SEM Characterization of fabricated plasmonic multispectral filters at various scale bars. (a) Overview image at 5 µm scale (b) Band – 1 at 2 µm scale (c) Band – 2 at 3 µm scale (d) Band – 3 at 3 µm scale (e) Band – 4 at 4 µm scale (f) Vertical cross-section of the filter after Ge deposition.

## 3. Detailed view of the thermal multispectral wheel

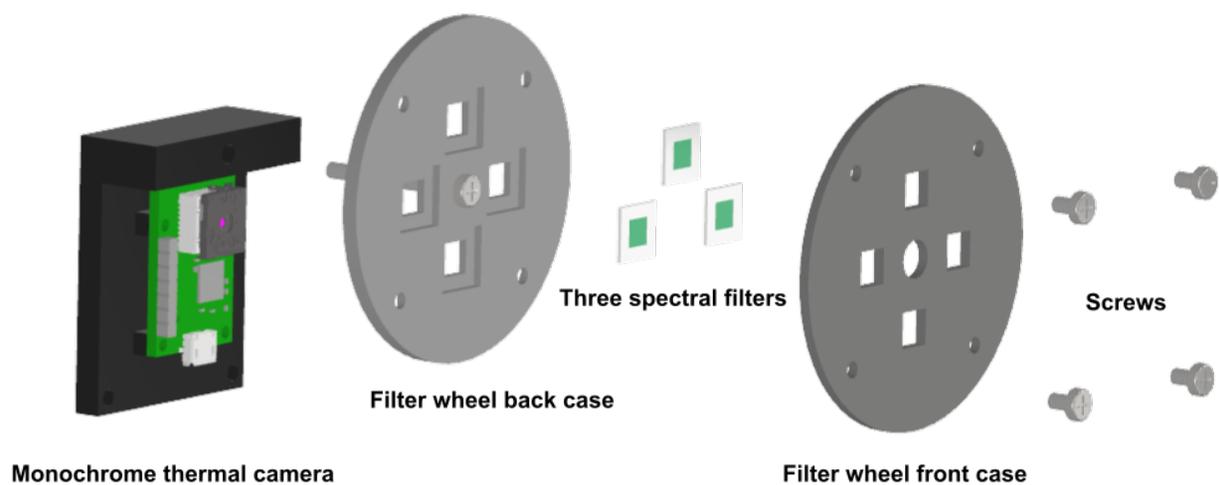

**Figure S3** Exploded schematic view of the thermal multispectral wheel. From left to right, monochrome thermal camera, filter wheel back case, three spectral filters, filter wheel front case and screws.

## 4. Superresolution network architecture

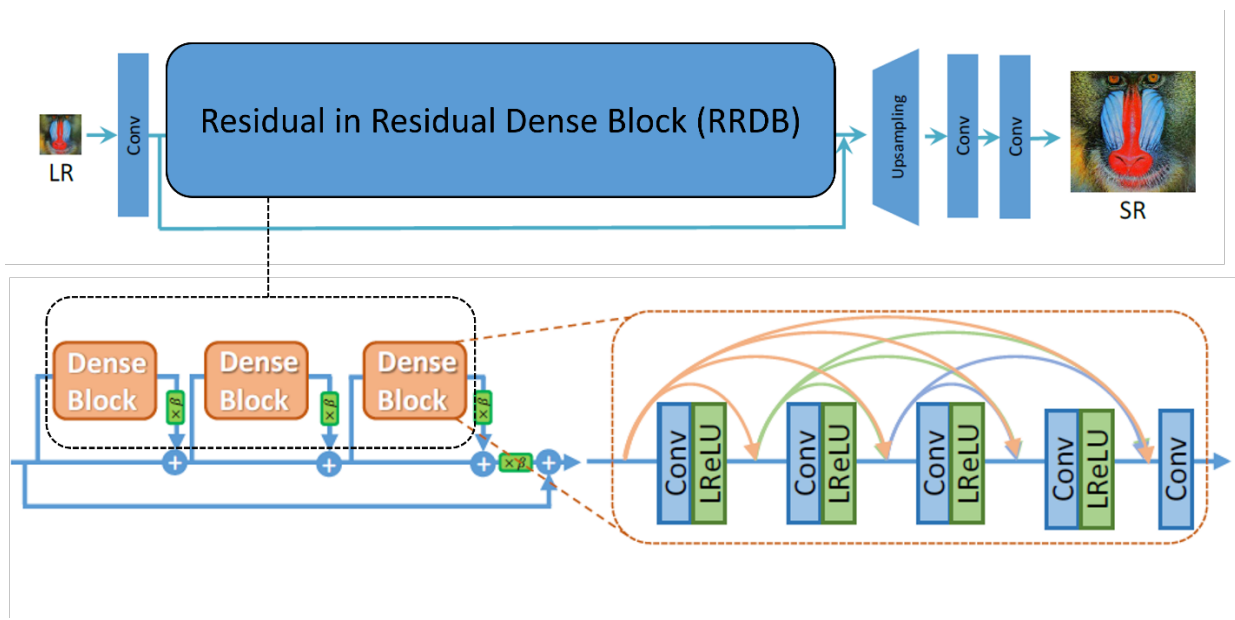

**Figure S4** Enhanced Superresolution Generative Adversarial Network (ESRGAN) Network Architecture for image superresolution [44]. RRDB block is adopted in the deep learning model and β is the residual scaling parameter.

## 5. Test Dataset

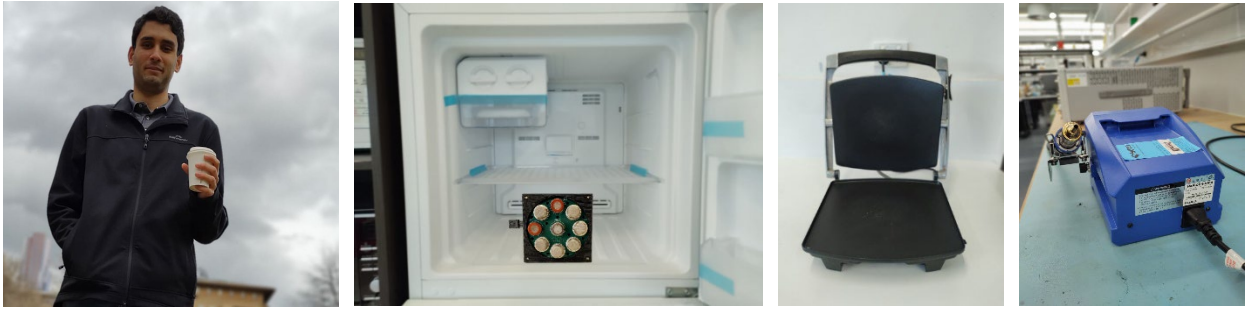

**Figure S5** From left to right: human holding a coffee (30 ºC), electronic nose (40 ºC), hot toaster (150 ºC) and soldering gun (200 ºC).

## 6. Test Dataset

| Parameter | Proposed System | Telops Camera [46] | Bodkin Camera [47] |
|---|---|---|---|
| *Detector Type* | Uncooled bolometer | Cooled InSb or MCT | Cooled |
| *Methodology* | Filter-wheel | Filter-wheel | Multiplexing |
| *Number of Bands* | 3 | 8 | 62 |
| *Spatial Resolution* | 80×60 (Software: 400 × 300) | 640×512 | 19×15 |
| *Spectral Range* | 8-14 µm | 7.5-11.5 µm | 7.8-10.8 µm |
| *Weight* | ~40 gms | ~13 kgs | ~25 kgs |
| *Packaging Size* | 77×62×45 mm | 321×199×176 mm | 406×330×914 mm |

**Table S1** Comparing the parameters of the proposed system in this work to commercial solutions, [46] and [47]